# NONLOCALITY IN QUANTUM THEORY

**Alexander V. Belinsky, Andrey K. Zhukovsky**

*Physics Department, Lomonosov Moscow State University*

This article describes a proposed way to conduct an experiment using a correlated particle pair in an entangled state, which leaves no room for any local models.

Contrary to the usual classical measurements, quantum measurements have a characteristic property that a physical quantity does not have any specific value if it does not occur in its eigenstate of measured quantity before the measurements take place (*a priori*) (*e.g.*, see [1] and references cited there). It is this property specifically and not the statistical character of measurements that singles out quantum theory as a separate branch of modern physics. Otherwise, it would be just a sub-branch of statistical physics. And it is this property specifically that fully conforms to the Copenhagen interpretation of quantum theory.

According to the projection postulate of von Neumann [2], during the measurement the state vector collapses, *i.e.*, its vector length is reduced to the measured range of values of the measured quantity (*e.g.*, see [1,3] and references cited there). Of special interest is usually the quantum state collapse of a pair or more of correlated particles being in entangled state, as the measurement of one particle leads to instantaneous quantum state change of the other (or others) that have moved away to an arbitrary, and, possibly, a significant distance from the first one. Strictly speaking, the von Neumann's projection postulate does not describe such a collapse. F.Ya.Khalili proposed a generalization of the postulate that would consider entangled states [4]. But the principal instantaneous collapse conclusion holds true. That's why we still see the attempts to create supraluminal communication lines based on that phenomenon (*e.g.*, see [5] and references cited there). But we think that detailed research of this phenomenon unquestionably testifies to the nonlocal character of quantum processes.

Consider the following experiment (see Fig.). Let's take two photons in the entangled state that are in anti-correlation state of polarization:

$$|\psi\rangle = \frac{1}{\sqrt{2}}(|0\rangle_x^a |1\rangle_x^b |1\rangle_y^a |0\rangle_y^b + |1\rangle_x^a |0\rangle_x^b |0\rangle_y^a |1\rangle_y^b). \tag{1}$$

The *a* and *b* indices here refer to the first and second photons of the entangled pair, and mutually orthogonal transverse directions of *x* and *y* determine the orthogonal polarization directions. This state vector has such a structure that even though the polarization directions *x* and *y* of each *a* and *b* pair photons are equally probable, they are closely linked between each other, or rather anti-correlated, as their polarization planes are mutually orthogonal.

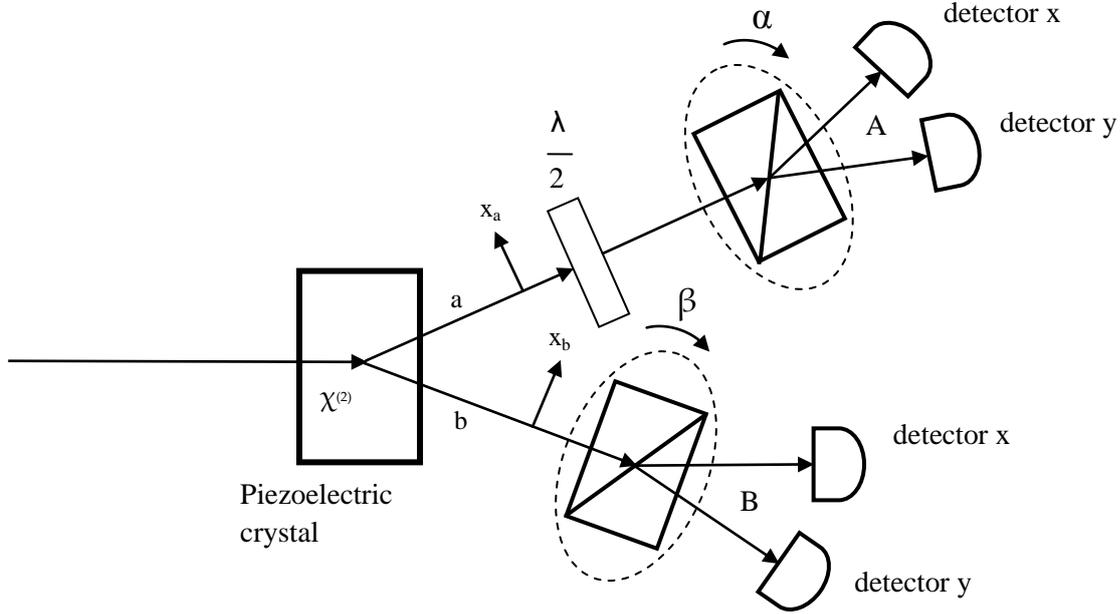

Fig.1. Research diagram for correlated photon pairs. Laser pumping in a piezoelectric crystal generates photon pairs. Signal *(a)* and idler *(b)* photons have mutually orthogonal polarization planes. They are directed, respectively, to observers *A* and *B* that both have Wollaston prisms, which separate mutually orthogonal polarizations, and two detectors: *x* and *y*. The observers have similar angular prism orientation, and it is determined, respectively, by $\alpha = \beta$ angles of turn around the photon path directions. Channel A has phase half-wave polarization plate installed. Polarization planes of its ordinary and extraordinary rays are oriented at an angle of π/4 to the respective planes of the piezoelectric crystal, i.e. it turns the photon's polarization plane by an angle of π/2.

Such states are usually prepared using parametric light scattering with type II non-linear interaction (*e.g.*, see [6] and references cited there).

Let's install phase half-wave polarization plate in one channel, say in channel *a*. The plate should be oriented in such a way that the polarization planes of its ordinary and extraordinary rays are at an angle of π/4 to the respective piezoelectric crystal planes. The plate is used to turn the polarization plane by π/2. Therefore, phase plate will change the state vector (1) in the following manner:

$$|\psi\rangle = \frac{1}{\sqrt{2}}(|1\rangle_x^a|1\rangle_x^b|0\rangle_y^a|0\rangle_y^b + |0\rangle_x^a|0\rangle_x^b|1\rangle_y^a|1\rangle_y^b). \qquad (2)$$

After photon *a* passes this λ/2 plate, make the measurement in channel *b*. Let's say that, for example, photon *b* after Wollaston prism finds itself in the channel with *y* polarization. In this case, after the measurement, the vector (2) collapses to

$$|\psi\rangle = \frac{1}{\sqrt{2}}(|1\rangle_y^a + |1\rangle_y^b), \qquad (3)$$

or just to $|1\rangle_y^a$, and photon *a* finds itself in *y* polarization channel after the polarization prism.

If, on the contrary, as a result of the measurement, photon *b* will find itself in the *x* polarization channel, everything will be vice versa.

There is no doubt that a real experiment will confirm this simple reasoning.

Now, let's try to proceed from such seemingly firmly established fact that *a priory* there is no measured quantity of the quantum observed value (*e.g.*, see [7]). In this case, before the measurement both photons have no polarization, *i.e.*, in best-case scenario – it is a natural polarization. The λ/2 plate can in no way impact such a photon, therefore, after photon *b* polarization state is measured, photon *a* should have acquired mutually orthogonal polarization, *i.e.*, the result of the experiment will be completely opposite.

Moreover, we could control the behavior of photon *a* by changing the photon *b* registration moment. Indeed, by increasing or decreasing the distance from nonlinear crystal to the Wollaston prism in channel *b*, we can control $|\psi\rangle$ state vector collapse moment: if this happens before photon *a* reaches the half-wave polarization plate, then both photons will register similar polarization states, and if this happens after, then according to the above reasoning, they would be opposite. But, as it was shown before, this of course will not happen. So, what does it mean?

It means that either the measured quantities have defined values before the measurement nonetheless, which would contradict not only the Copenhagen interpretation but the results of other experiments as well, *e.g.*, [7]. Or, we should draw a conclusion that although measured quantities do not have *a priori* values, they do have strict correlation between them, *i.e.* we are observing a kind of a correlation of non-existent things!

Both alternatives are amazing and leave no room for simple local models. In the first case we would have to talk about nonlocal theory of hidden parameters, because only this theory can explain the disruption of Bell inequalities, which was found in a number of experiments [8,9], as well as many others. And in the second, we will have to talk about nonlocal correlation of physical quantities that have no specific values.